\documentclass[fleqn,usenatbib]{mnras}
\usepackage{newtxtext,newtxmath}
\usepackage[T1]{fontenc}
\usepackage{ae,aecompl}
\usepackage{amsmath}
\usepackage{amssymb}
\usepackage{ulem}
\usepackage{graphicx}
\usepackage{times}
\usepackage{hyperref}

\newcommand{\kepler}{\textit{Kepler}}

\newcommand{\Edit}[1]{#1}
\newcommand{\Move}[1]{#1}
\newcommand{\EditTwo}[1]{#1}
\newcommand{\Nsample}{60}
\newcommand{\Nfixed}{15}
\newcommand{\Nfixedpercent}{25\%}
\newcommand{\NfixedTwo}{6}
\newcommand{\NfixedTwoOneSeven}{9}
\newcommand{\Nfixedwjitter}{5}
\newcommand{\NfixedwjitterTwoOneSeven}{4}

\newcommand{\Nextendedsample}{95}
\newcommand{\Nfixedextended}{30}
\newcommand{\Nfixedextendedpercent}{31\%}
\newcommand{\NfixedTwoOneSevenextended}{22}
\newcommand{\NfixedTwoextended}{8}

\newcommand{\Nfixedfloat}{19}
\newcommand{\Nfixedfloatpercent}{32\%}
\newcommand{\Nfixedfloatextended}{41}
\newcommand{\Nfixedfloatextendedpercent}{43\%}
\newcommand{\NfixedfloatTwo}{4}
\newcommand{\NfixedfloatTwoOneSeven}{6}
\newcommand{\NfixedfloatFloating}{9}
\newcommand{\Nfixedfloatwjitter}{4}

\title[Mischaracterization of exoplanetary systems]{Systematic mischaracterization of exoplanetary system dynamical histories from a model degeneracy near mean-motion resonance}

\author[J. H. Boisvert et al.]{
John H. Boisvert,$^{1}$
Benjamin E. Nelson,$^{2}$
and Jason H. Steffen$^{1}$\thanks{E-mail: jason.steffen@unlv.edu}
\\
$^{1}$Department of Physics and Astronomy, University of Nevada, Las Vegas,4505 S Maryland Parkway, Box 454002, Las Vegas, NV 89154-4002, USA\\
$^{2}$CIERA -- Northwestern University, 2145 Sheridan Road, Evanston, IL 60208-3112, USA\\
}
\date{Accepted XXX. Received YYY; in original form ZZZ}
\pubyear{2018}

\begin{document}
\label{firstpage}
\pagerange{\pageref{firstpage}--\pageref{lastpage}}
\maketitle
\begin{abstract}
There is a degeneracy in the radial velocity exoplanet signal between a single planet on an eccentric orbit and a two-planet system with a period ratio of 2:1.  This degeneracy could lead to misunderstandings of the dynamical histories of planetary systems as well as measurements of planetary abundances if the correct architecture is not established.  We constrain the rate of mischaracterization by analyzing a sample of \Nsample\ non-transiting, radial velocity systems orbiting main sequence stars from the NASA Exoplanet Archive (NASA Archive) using a new Bayesian model comparison pipeline.  We find that \Nfixed\ systems (\Nfixedpercent\ of our sample) show compelling evidence for the two-planet case with a confidence level of 95\%.
\end{abstract}


\section{Introduction} \label{sec:intro}

The architectures of planetary systems give insight into their formation and dynamical histories.  For example, interactions with the protoplanetary disk tend to drive adjacent planets into first-order, mean-motion resonances (MMRs, such as the 2:1), while simultaneously damping their eccentricities to values that are difficult to measure \citep{LeePeale02,Tinney06}.  On the other hand, planet-planet scattering \citep{Chatterjee08,FordRasio08} or Kozai-Lidov oscillations \citep{Kozai62,Lidov62,Fabrycky07} can produce single planets with eccentric orbits.  While not all planetary systems must pass through these phases of disk migration or eccentricity growth, the system architectures that they produce rarely occur from in-situ formation.  Thus, reliable estimates of their frequencies will reveal the relative importance of these processes in planet formation and evolution in general.

For radial velocity (RV) observations in particular, the challenge in identifying the true system architecture is a degeneracy between two models---one with a single planet with eccentric orbits (single eccentrics) and one with two planets with circular orbits at the 2:1 (circular doubles) \citep{Angelada10,Wittenmyer13}.  Historically, the single-planet model has been favored on the grounds of Occam's razor \citep{Kurster15}, since a system with a single planet is simpler than a system with two.  However, the circular double model has the same number of model parameters as single eccentrics (it is just as simple) and it is a consequence of dynamical processes known to occur.  These facts motivate careful scrutiny of existing discoveries in order to properly characterize the systems.  If circular doubles are more common than currently suggested, then disk-migration may be more important than previously thought \citep{Tinney06}.

The source of the degeneracy between these models is in a first-order expansion of the RV signal of a single eccentric planet
\begin{equation}
RV_{\mathrm{single}} \approx K \cos (M + \omega) + K e \cos (2M + \omega) + \mathcal{O}(e^2),
\label{eq:SingleKeplerian}
\end{equation}
where $RV_{\mathrm{single}}$ is the observed radial velocity, $K$ is the velocity semi-amplitude, $e$ is the eccentricity, $\omega$ is the longitude of periastron, and $M$ is the mean anomaly\Edit{, which is a function of time}.  By comparison, the signal of a circular double is
\begin{equation}
RV_{\mathrm{double}} = K_{\mathrm{out}} \cos (M_{\mathrm{out}}) + K_{\mathrm{in}} \cos(M_{\mathrm{in}}),
\label{eq:CircularDouble}
\end{equation}  
where $RV_{\mathrm{double}}$ is the observed radial velocity, $K_{\mathrm{out}}$ and $K_{\mathrm{in}}$ are the velocity semi-amplitudes, and $M_{\mathrm{out}}$ and $M_{\mathrm{in}}$ are the mean anomalies.  At the 2:1 MMR, $M_{\mathrm{in}}=2 M_{\mathrm{out}}$ and the inner planet signal ($K_{\mathrm{in}}$) masquerades as the eccentricity signal ($K e$) of the single planet.

This degeneracy is widely known though rarely addressed.  Nevertheless, there is precedent for reconsidering certain systems.  For example, \citet{Kurster15} reanalyzed RV data for HD~27894 and found that a circular double model was a better fit than the reported single eccentric model.  Also, \citet{Angelada10, Wittenmyer13}, found similar results for several RV systems.  At the same time, new measurements from the \kepler\ mission show that planet pairs near 2:1 are quite common.  For example, using the method of \citet{Steffen15} on the \textit{Kepler} DR25 catalog \citep{Thompson17}, we estimate that 20$\%$ of \textit{Kepler}'s transiting adjacent planet pairs with period ratios between 1 and 6 are within ten percent of 2---including the most prominent peak of the period ratio distribution at 2.17 \citep{Steffen15}.  Motivated by these new facts and the results of previous studies, we reanalyze a sample of \Nsample\, single eccentric planetary systems using a new Bayesian analysis pipeline in an effort to discover their true architectures.

\section{\Edit{Methods}}\label{sec:setup}
 
\begin{figure}
\includegraphics[width=\columnwidth]{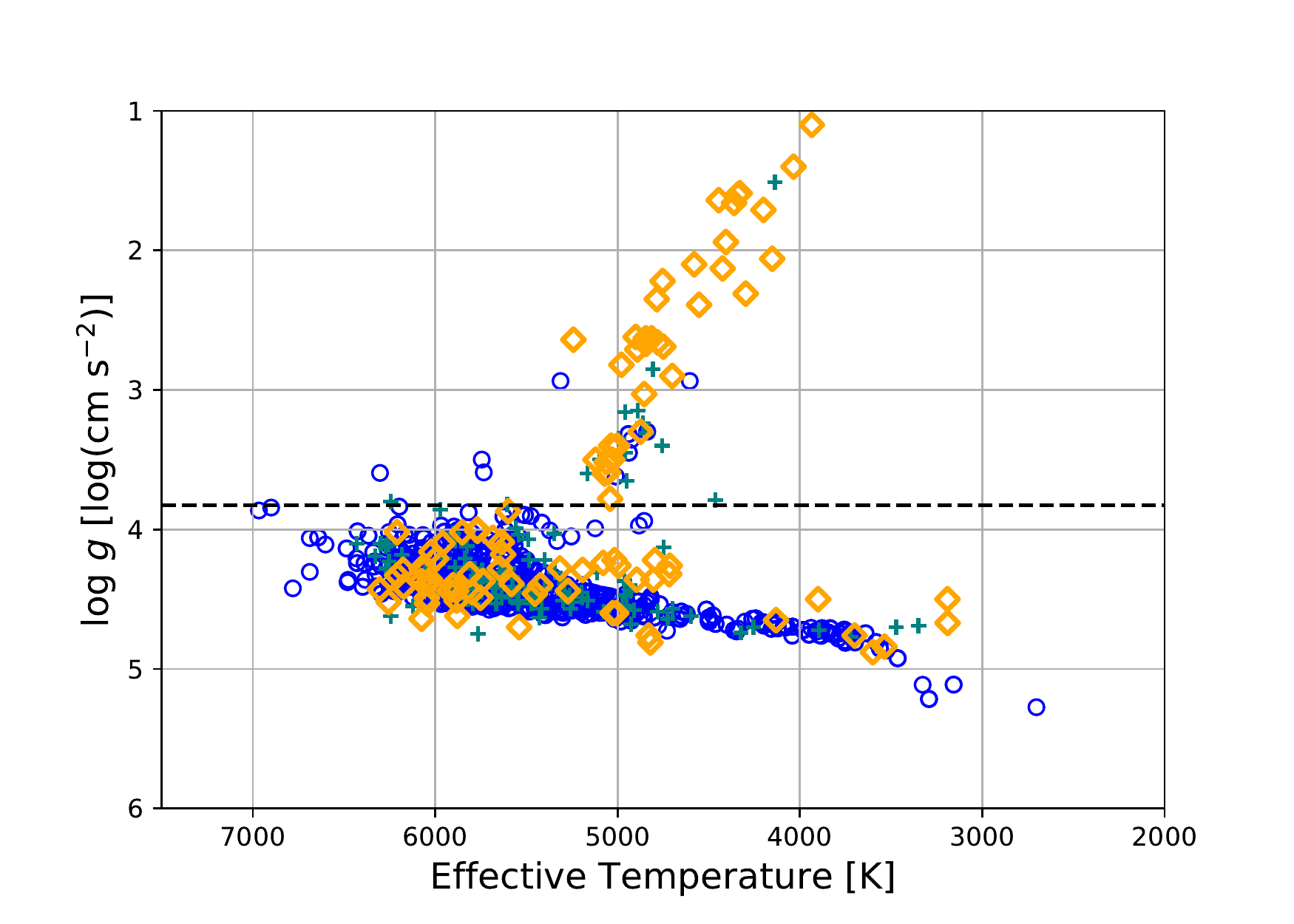}
\caption{Stellar effective temperature vs.$\,\,$stellar surface gravity for the RV multi-planet systems from the NASA Archive as grey crosses and the \kepler\ multi-planet systems as blue circles.  Our sample of 95 stars, ignoring stellar type, are orange diamonds.  For our sample of main sequence stars we selecte those with $\mathrm{log}\,g\geq3.825$, there are 60 main sequence stars in the main sample and 95 stars in the entire sample.}
\label{fig:astellarcomparison-entire}
\end{figure}

Our sample contains \Nsample\ systems and comprises every non-transiting RV system from the NASA Exoplanet Archive (NASA Archive), as of November 2016 \citep{Akeson13}, that is listed as having only a single planet, orbiting a main sequence star, and whose system properties were derived from a single data set.  \Edit{Our pipeline did not have the capability to analyze multiple datasets when the analysis began.  We did not limit our sample by eccentricity.}  Our main results will focus on the main sequence stars, but we will also report on an extended sample which ignores stellar type.  The extended sample contains \Nextendedsample\ systems, which is nearly a quarter of all RV-discovered single-planet systems.   Figure~\ref{fig:astellarcomparison-entire} shows how we determine stellar types based on their reported surface gravity and how we select the main sequence sample.

\subsection{\Edit{The Pipeline}}\label{sec:pipeline}
Our pipeline estimates the Bayes factor---the ratio of the probabilities of the RV data given the circular double model and the single eccentric model---to quantitatively compare the two models.  For each system we test four planetary system models: a single eccentric; two circular doubles (one with a period ratio fixed at 2 and the other fixed at 2.17---where there are two large peaks in the period ratio distribution from \kepler\ \citep{Steffen15}); and a ``floating'' circular double with no period ratio constraint.  This last model has an additional model parameter, but the Bayes factor calculation can account for different numbers of model parameters.  Our primary results work with the two fixed models given the compelling theoretical and observational reasons to consider them, the fact that the number of model parameters are identical (and thus more directly comparable), and because a narrow-band signal at a fixed period ratio is less susceptible to a false positive detection from stellar RV jitter or statistical noise.  

The parameters for the single eccentric model are: period ($P$), velocity semi-amplitude ($K$), eccentricity ($e$), longitude of periastron ($\omega$), and mean anomaly at the time of the earliest RV measurement ($M_\mathrm{o}\equiv M(t_\mathrm{o})$).  The parameters for the circular double models are: outer planet period ($P_{\mathrm{out}}$), the velocity semi-amplitude for the outer/inner planets ($K_{\mathrm{out}}$/$K_{\mathrm{in}}$), and mean anomaly at the time of the earliest RV measurement for the outer/inner planets ($M_{\mathrm{o,out}}$/$M_{\mathrm{o,in}}$). The floating circular double model includes the inner planet orbital period ($P_{\mathrm{in}}$).  Each model also has a linear trend ($A$) and a velocity offset ($C$).   

Our model fitting is a three step process.  First, we determine the starting values for our Markov Chain-Monte Carlo (MCMC) by maximizing the likelihood function:
\begin{equation}
\mathrm{ln} \; P(t,RV,\sigma_{RV}|\theta) = -\frac{1}{2} \sum_{n} \left [ \frac{(RV_n - \mathcal{M}(\theta,t_n))^2}{\sigma_{RV,n}^2} + \mathrm{ln} \; 2\pi\, \sigma_{RV,n}^2 \right ]
\label{eq:likelihood}
\end{equation}
where $t$, $RV$, and $\sigma_{RV}$ are the observed times, RV measurements, and RV errors.  $\mathcal{M}$ is the RV model and $\theta$ are its associated parameters.  We draw our set of initial conditions for the maximum likelihood estimation (MLE) from the NASA Archive.  \Edit{The time of periastron passage is used to determine the initial $M_{\mathrm{o}}$.}  We first fit for $C$, fixing the other parameters at their nominal values and setting $A=0$.  We next fit for $A$ and $C$ simultaneously.  Some systems did not have $K$, $\omega$, and/or \Edit{the time of periastron passage} reported on the NASA Archive.  In those cases, an MLE was done with the missing quantities as the only free parameters.  We initialized the fixed circular double models to their first-order, single eccentric equivalent values using equations (\ref{eq:SingleKeplerian}) and (\ref{eq:CircularDouble}).  For the floating circular double, the inner planet orbital period is initialized to either the 2:1 or the 2.17:1, depending on which fixed model produced a larger Bayes factor.

The second step in our pipeline estimates the posterior distributions of the model parameters using an ensemble sampler MCMC \citep{DFM13}.  Each run has thirty Markov chains, thins the chains every hundred steps, and ignores the first 20\% of the chain as burn-in. We allow the chains to evolve until they yield a set of at least ten thousand independent samples per model per RV dataset.  \Edit{We measure the autocorrelation length after each run to determine the number of independent samples.  If the number of independent samples falls short of ten thousand, then the autocorrelation length is used to determine how many additional steps are needed to yield ten thousand independent samples and the \EditTwo{MCMC} is rerun with the new number of steps.}  The different chains were initialized using the parameter values from the MLE, with each parameter scattered by a sufficiently small amount to allow the ensemble sampler to fill the posterior mode.  

\Move{We impose a modified Jeffery's prior for the orbital period and velocity semi-amplitude: $p(X)=[(1+X)\times\mathrm{ln}(1+X_{\mathrm{max}}/X_0)]^{-1}$ with bounds between 0--10,000~days and 0--2,000~m~s$^{-1}$ respectively and $X_0$ equal to 1 day and 1 m s$^{-1}$ respectively.  We use this prior because it is normalizable, objective, and intended for scalable parameters that could have zero as a value.  We use uniform priors for the remaining parameters, ($e$, $\omega$, $M_0$), because they are also normalizable and objective.  We sample the parameters for the single eccentric model in \{$P$, $K$, $\sqrt{e} \sin (\omega)$, $\sqrt{e} \cos (\omega)$, $\omega+M_0$\}-space in order to maintain uniform priors \citep{DFMBen}. 

The prior bounds for $K$, $K_{\mathrm{out}}$, and $K_{\mathrm{in}}$ are between 0 and 2,000 m s$^{-1}$.  The prior bounds for $P$, $P_{\mathrm{out}}$, and $P_{\mathrm{in}}$ are between 0 and 10,000 days. The prior bounds for $\sqrt{e} \, \sin (\omega)$ and $\sqrt{e} \cos (\omega)$ are such that $0 < (\sqrt{e} \sin (\omega))^2 + (\sqrt{e} \cos (\omega))^2<1$, i.e.~$0<e<1$.  The prior bounds for $(M_0+\omega)$, $M_{\mathrm{0,out}}$, and $M_{\mathrm{0,in}}$ are between -2$\pi$ and 4$\pi$.  These limits allow the Markov chains to cross the 0 and 2$\pi$ coordinate singularities while remaining well-behaved.
Furthermore, these values are modded by 2$\pi$ before doing any calculations.
The prior bounds for $C$ are between -100,000 m s$^{-1}$ and 100,000 m s$^{-1}$ to accommodate the wide range in offset values in the real sample.}

Finally, we estimate the Bayes factors between the single eccentric model and the circular double models by taking the ratio of the fully marginalized likelihoods (FML, i.e. Bayesian evidence) for the two models.  We approximate the FML using an importance sampling algorithm where the sampling distribution is informed by a set of posterior samples taken from the aforementioned MCMC \citep{Nelson16}.  For each system we take the larger of the Bayes factor for the two fixed circular double models.

\Move{In this context, importance sampling is essentially a general form of Monte Carlo integration to estimate the fully marginalized likelihood, $\mathcal{Z}$.
The value of $\mathcal{Z}$ is the integral over the prior probability distribution $p(\theta)$ times the likelihood function $\mathcal{L}(\theta) \equiv p(t, RV, \sigma_{RV}|\theta)$, i.e., 
\begin{equation}
\mathcal{Z} = \int p(\theta)\mathcal{L}(\theta)d\theta
\end{equation}
We multiply the numerator and denominator of the integrand by $g(\theta)$, a distribution over the model parameters with a known normalization.
\begin{equation}
\mathcal{Z}=\int \frac{\mathcal{L}(\theta)p(\theta)}{g(\theta)} g(\theta) d\theta.
\label{eq-mcmcis-1}
\end{equation}
Equation \ref{eq-mcmcis-1} is in a form such that $\mathcal{Z}$ can be estimated numerically by drawing $N$ samples from $g(\vec{\theta})$, 
\begin{equation}
\widehat{\mathcal{Z}} \approx \frac{1}{N}\sum\limits_{\theta_i \sim g(\theta)}\frac{\mathcal{L}(\theta_i)p(\theta_i)}{g(\theta_i)}.
\label{eq-mcmcis-2}
\end{equation} 

The key to an accurate and efficient estimate of $\widehat{\mathcal{Z}}$ lies in choosing an appropriate $g(\theta)$.
Assuming our parameter space contains one dominant posterior mode, we choose a multivariate normal $\mathcal{N}(\vec{\mu}_g, \vec{\Sigma}_g)$, where $\vec{\mu}_g$ and $\vec{\Sigma}_g$ describe the mean vector and covariance matrix of the model parameters respectively.
After we perform an MCMC on a particular model/dataset, we can estimate $\vec{\mu}_g$ and $\vec{\Sigma}_g$ using a set of posterior samples.  That information is fed into our importance sampling algorithm to estimate $\widehat{\mathcal{Z}}$ for that model.
\citet{Nelson16}, \citet{Guo12}, and \citet{Weinberg13} provide more detailed prescriptions and investigations of this method.}

\subsection{\Edit{Pipeline Characterization}}\label{sec:pipelinecharacter}
\begin{figure}
\includegraphics[width=\columnwidth]{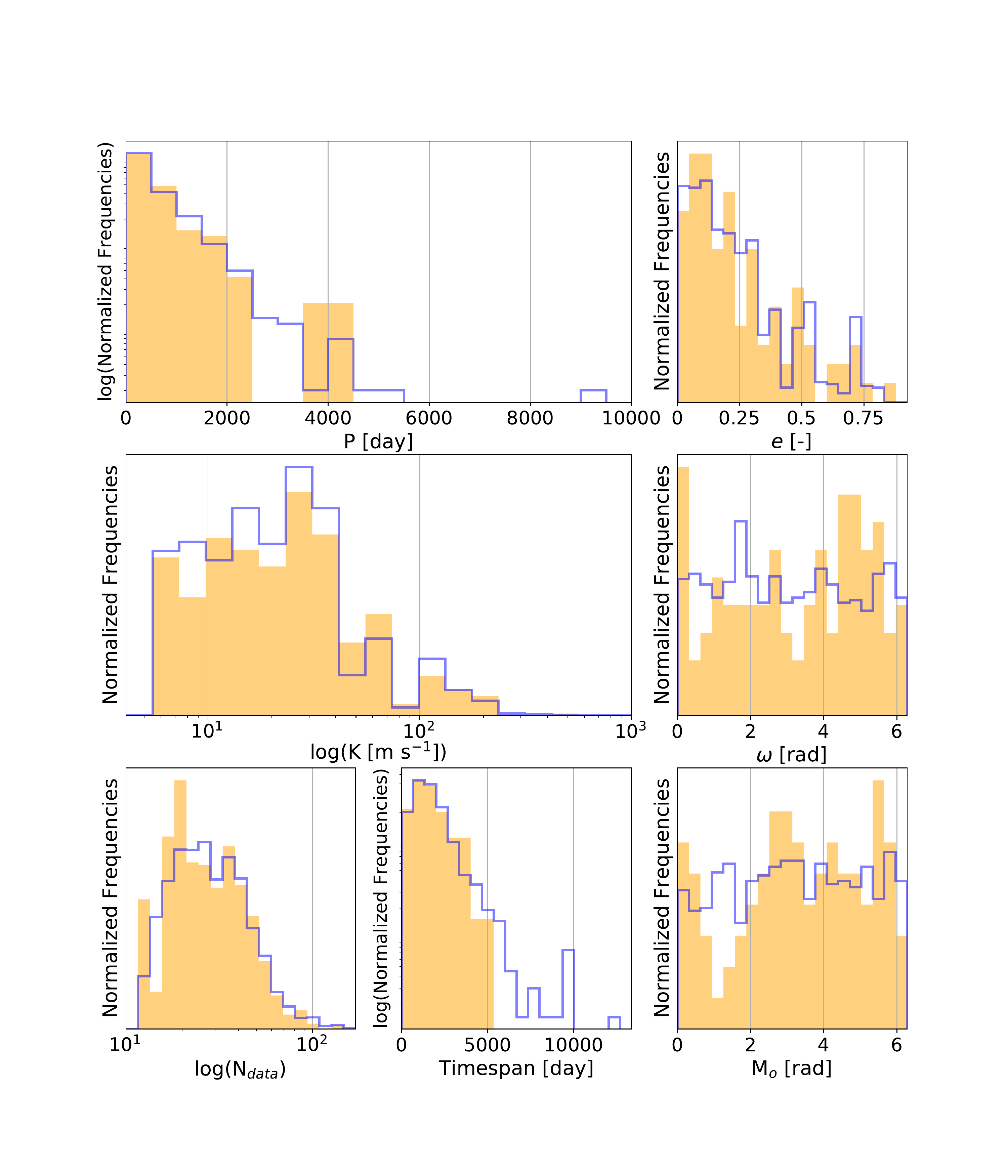}
\caption{Property distributions for our sample of 95 real systems from the NASA Exoplanet Archive in orange and the 1,000 synthetic time series in blue.}
\label{fig:parametercomparison}
\end{figure}

We characterized the pipeline efficiency with an ensemble of 1,000 synthetic RV time series whose system and data properties match the real systems.  \Move{We use the Bayes factors of these synthetic systems to characterize our model comparison pipeline.  This Monte Carlo simulation was initialized as follows.

The start time ($t_0$) is \Edit{a uniform} random \Edit{draw} between 1 and 1,000 days.  The number of observations are drawn from the real systems with a normally-distributed adjustment with a standard deviation 10\% of the nominal value rounded to the nearest whole number.  
The observation time series is produced by selecting a set of observation differences ($t_{i}-t_{i-1})$ from the real distribution of observation differences with a similar, normally-distributed 10\% variation added to each difference.  The number of orbits is the number of orbits of a randomly chosen real system with a normally distributed 10\% variation.

We determine the orbital period ($P$) using the selected number of orbits and the observation time series.
The velocity semi-amplitude ($K$) and the eccentricity ($e$) are separate random draws from the real systems.  The mean anomaly of the start time ($M_0$) and argument of periastron ($\omega$) are randomly drawn between 0 and $2\pi$. 
The linear trend ($A$) is a 10\% variation to a random draw from the real systems. 

We assume that the RV errors are normally distributed with a standard deviation that is the quadrature sum of stellar jitter and instrumental and photon noise.  The instrumental and photon noise ($\sigma_{\text{RV}}$) are drawn randomly from the RV errors of the real systems and our error bars are assigned to this value.  Steller jitter is selected from a log uniform distribution between 0.5 and 5 m s$^{-1}$.   
The observation errors are added to the synthetic RV measurement---not to the error in the RV measurement.  Figure~\ref{fig:parametercomparison} shows the parameter distributions for the 1,000 synthetic time series and the real systems as reported in the NASA Archive.}

\begin{figure}
\includegraphics[width=\columnwidth]{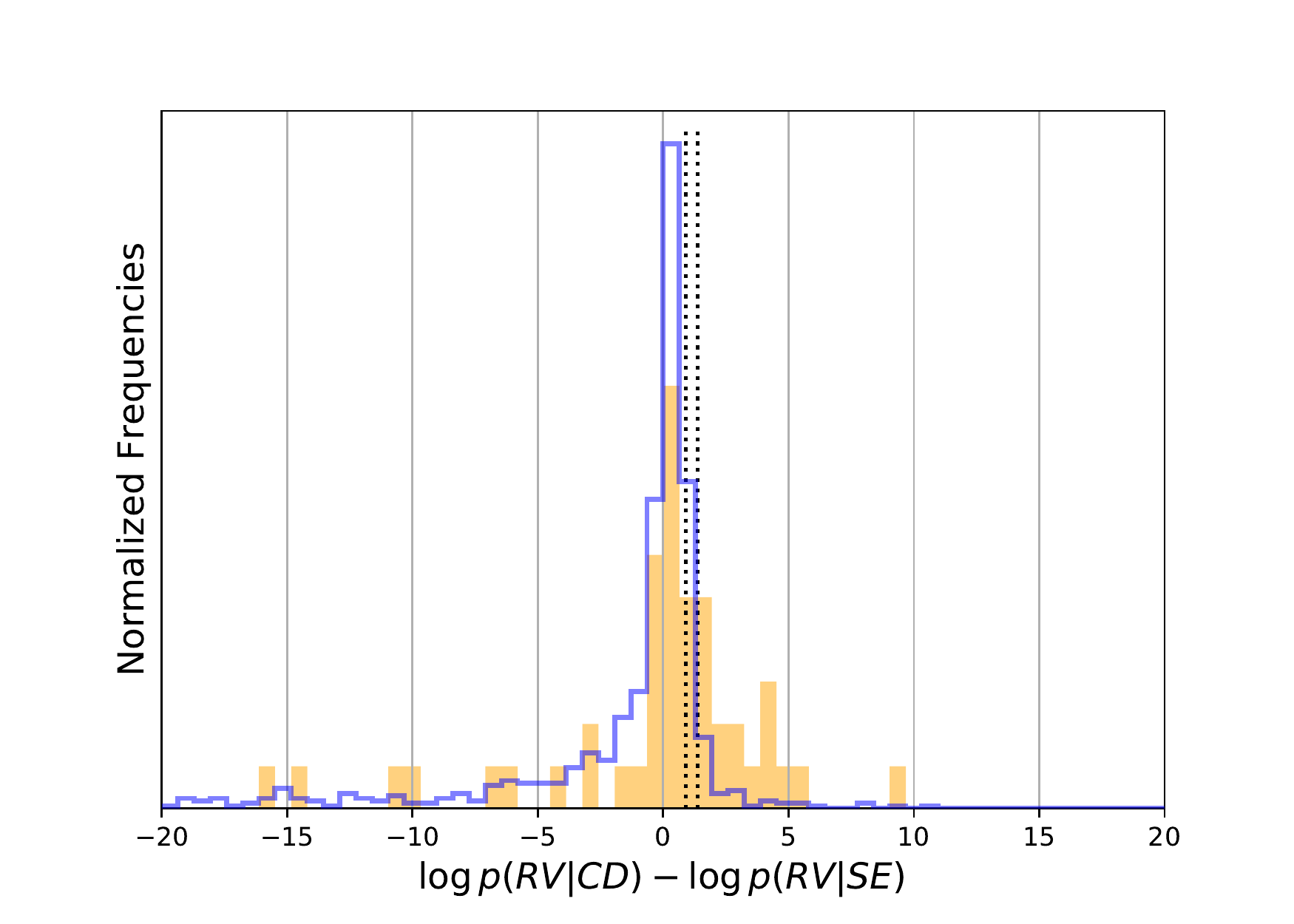}
\caption{The log Bayes factor distribution for the 1,000 synthetic single eccentric time series in blue and 60 real systems hosted by main sequence stars in orange.  Here, we compare only the single eccentric model to the fixed, circular double model with the largest Bayes factor.  The 95th and 90th percentiles are indicated with the dotted lines near a Bayes factor of 24 and 8, respectively.
}
\label{fig:bayesrealvssynthMS}
\end{figure}

The resulting Bayes factors from this characterization are shown as the blue distribution in Figure~\ref{fig:bayesrealvssynthMS}.  The vertical lines denote the 95th and 90th percentiles of the distribution. The shape of the distribution is not symmetric, and the vast majority of our synthetic datasets favor the single-planet case---as expected since the synthetic systems were constructed to be single eccentrics.  
Real systems with Bayes factors larger than those thresholds may be circular double systems mischaracterized as single eccentrics. 

The approach outlined above is different from earlier studies.  For example, \citet{Wittenmyer13} used the reduced $\chi^2$ to determine the preferred model and refined their results with stability tests using the N-body integrator \textit{Mercury} \citep{Chambers97}.  \citet{Angelada10} randomized individual sets of data to calculate the false positive rate per system.  Their model selection was also based on the reduced $\chi^2$ of least-squares fitting.  In this work, we use a fully marginalized likelihood to calculate the Bayes factor for the model comparison and we estimate our false positive rate by analyzing a large simulated dataset with our pipeline.
 
\section{Results}\label{sec:results}

After analyzing our synthetic systems, we ran our sample of \Nsample\ real systems (\Nextendedsample\ systems for the extended sample) through the same pipeline.
Figure~\ref{fig:bayesrealvssynthMS} shows that the Bayes factor distributions for the synthetic and real systems (in orange) are not similar.  We find that \Nfixed\ (\Nfixedpercent)  of the systems have Bayes factors larger than the 95th percentile of the synthetic systems.  (For the extended sample of \Nextendedsample\ systems the numbers are \Nfixedextended\ and \Nfixedextendedpercent\ of the entire sample respectively.)  \NfixedTwoOneSeven\ of these systems prefer the 2.17:1 model (\NfixedTwoOneSevenextended\ of the extended sample) while the remaining \NfixedTwo\ (\NfixedTwoextended\ from the extended sample) prefer the 2:1 model.  Assuming a false positive rate of 5\% from our 95\% confidence level, our estimate of the number of false positives is 0.75 $\pm$ 0.87 (1.5 $\pm$ 1.2 for the extended sample).  \Edit{The systems from the extended sample that prefer the fixed circular double model, the model parameters, and Bayes factors are shown in Table~\ref{tab:Results}.}  A CSV file containing the model parameters with errors for all four models, Bayes factors between the circular double models and the single eccentric model, and percentile of the best fixed model for each system in the extended sample are available online as \Edit{Table~2}.

\begin{figure}
\includegraphics[width=\columnwidth]{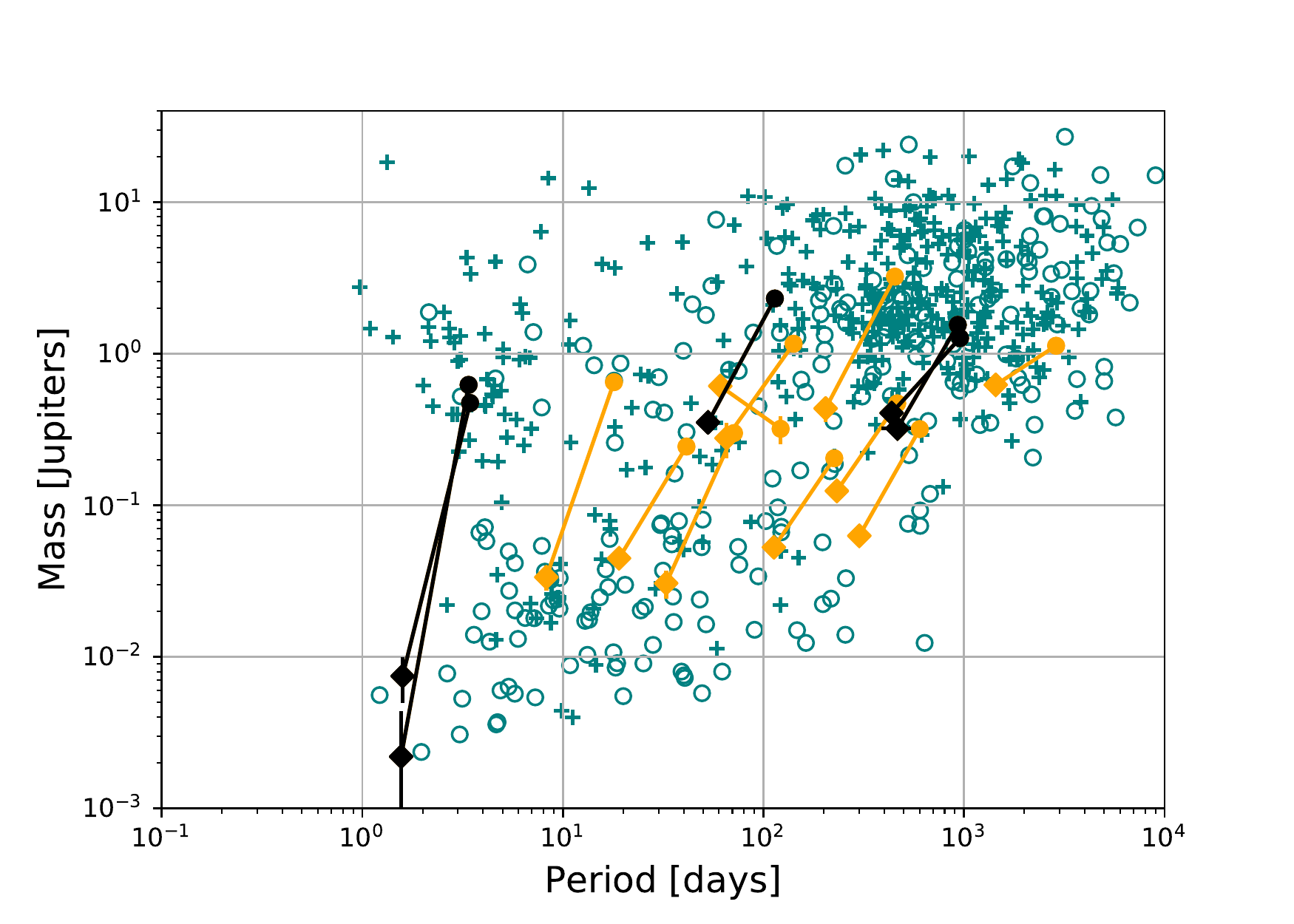}
\caption{Orbital period vs.$\,\,$planetary mass for all RV planets.  Systems from the NASA Exoplanet Archive are in teal, with multi- and single-planet systems as open circles and crosses, respectively.  Each system orbiting a main sequence star with a measured Bayes factor larger than 95th percentile of the synthetic systems are plotted in orange.  Each putative system is represented by a line on the plot, with the diamonds as the inner planet and the circles are the outer companion.  Systems that remain in the 95th percentile after including a white noise stellar jitter term are in black.  We note that these results lie well within normal parameter values of known systems.}
\label{fig:massvsperiodMS}
\end{figure}

\begin{figure}
\includegraphics[width=\columnwidth]{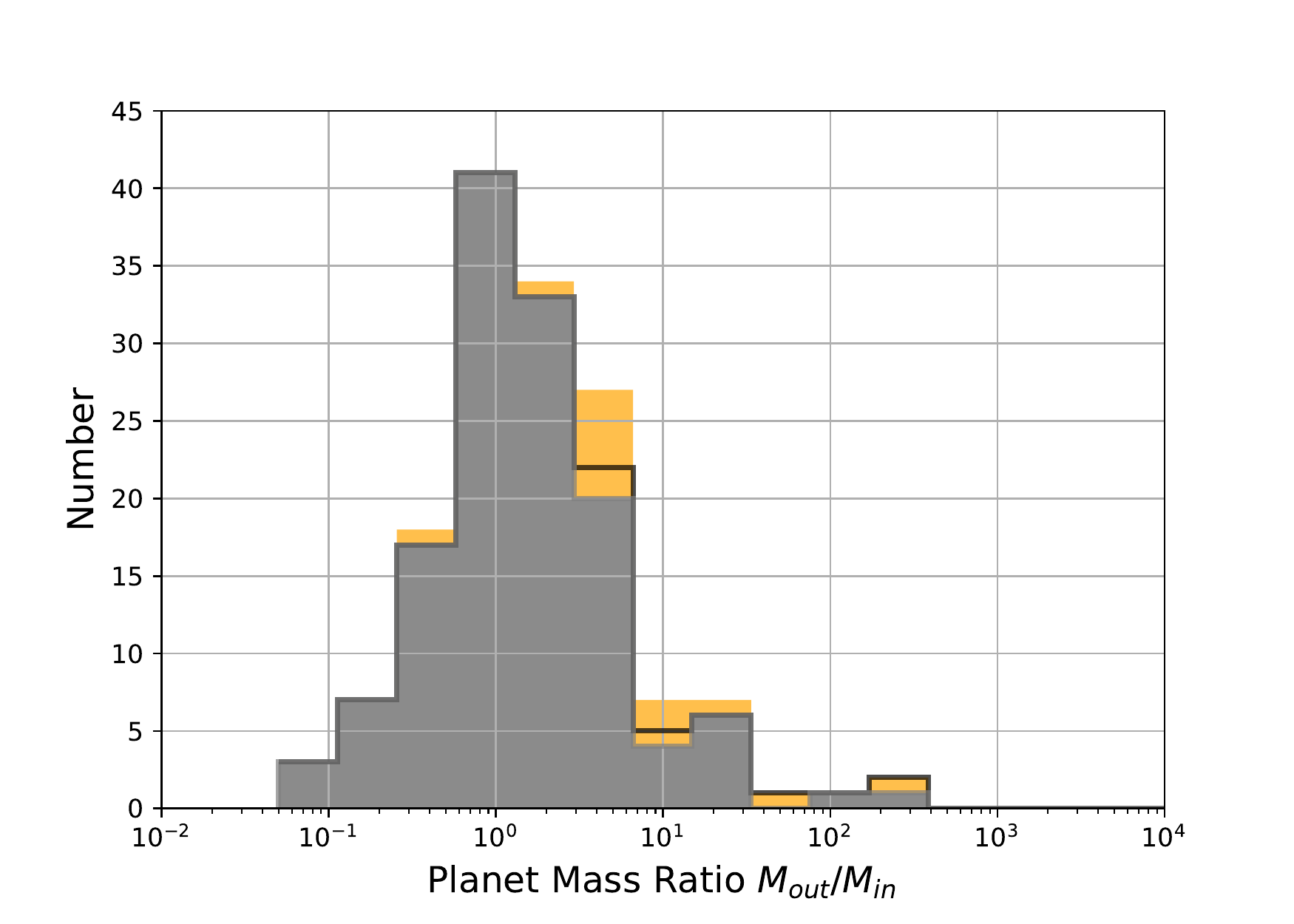}
\caption{Mass ratio distribution for all RV adjacent planet pairs in grey.  The stacked, orange distribution are our systems with Bayes factors larger than 24 (95th percentile) around main sequence stars.  The nature of the signal favors more massive outer planets.  The black outline shows the systems that have stellar jitter included and still had Bayes factors larger than the 95th percentile.}
\label{fig:PRHistogramPlusMS}
\end{figure}

We examine the consequences of these potential discoveries on several distributions of planet properties.
Figure~\ref{fig:massvsperiodMS} shows the planet mass vs.$\,\,$orbital period for known RV planets along with the new planets favored by our analysis orbiting main sequence stars.  These potential new systems lie well within the range of values measured in known systems.
We point out, however, that some systems may yet be false positives.  For instance, there are a few \Edit{candidate circular double} systems that \EditTwo{would be} hot Jupiters (planet with $P \lesssim 10$ days) with interior companions.  Presently, there is only a single known system (WASP-47 \citep{Becker15}) where a hot Jupiter has a known interior companion.  And the period ratio in this case is over 5:1---far from the degeneracy we consider here.  However, \Edit{the hot Jupiter has an} outer companion \Edit{with} a period ratio near 2.17.  Figure~\ref{fig:PRHistogramPlusMS} shows how the predicted mass ratios for the main sequence systems that favor the two-planet model compare with the mass ratios for RV systems on the NASA Archive. 

\onecolumn
\begin{table}
\begin{center}
\fontsize{7.5}{7.2}\selectfont

\rotatebox{90}{
\begin{tabular}{ |c|c|c|c|c|c|c|c|c|c|c|c| }
\hline
Star Name	&	Star Type	&	Model	&		$P_{\mathrm{out}}$ [days]		&		$K_{\mathrm{out}}$ [m s$^{-1}$]		&		$M_{\mathrm{out}}$ [rad]		&		$P_{\mathrm{in}}$ [days]		&		$K_{\mathrm{in}}$ [m s$^{-1}$]		&		$M_{\mathrm{in}}$ [rad]		&		A [\EditTwo{k}m s$^{-1}$\EditTwo{day$^{-1}$}]		&		C [m s$^{-1}$]		&	Bayes Factor	\\
\hline
\hline
HD 240237	&	nMS	&	2.17:1	&	770.45	$\pm$	3.09	&	69.75	$\pm$	1.59	&	5.45	$\pm$	0.04	&	355.05	$\pm$	3.09	&	57.52	$\pm$	2.57	&	1.43	$\pm$	0.08	&	9.96	$\pm$	2.49	&	82.31	$\pm$	1.54	&	1.09 $\times$ 10$^{85}$	\\
HD 2952	&	nMS	&	2.17:1	&	318.83	$\pm$	0.27	&	23.16	$\pm$	0.94	&	4.13	$\pm$	0.05	&	146.92	$\pm$	0.27	&	15.90	$\pm$	0.71	&	4.41	$\pm$	0.10	&	6.37	$\pm$	0.62	&	3.15	$\pm$	0.74	&	6.88 $\times$ 10$^{57}$	\\
$\alpha$ Arietis	&	nMS	&	2.17:1	&	372.60	$\pm$	0.52	&	29.67	$\pm$	0.52	&	4.31	$\pm$	0.04	&	171.70	$\pm$	0.52	&	10.25	$\pm$	0.46	&	6.17	$\pm$	0.10	&	-9.5	$\pm$	0.64	&	0.91	$\pm$	0.45	&	9.80 $\times$ 10$^{49}$	\\
HD 96127	&	nMS	&	2.17:1	&	632.39	$\pm$	2.19	&	96.66	$\pm$	1.50	&	5.95	$\pm$	0.04	&	291.42	$\pm$	2.19	&	41.02	$\pm$	1.26	&	1.41	$\pm$	0.07	&	22.83	$\pm$	3.49	&	-923.95	$\pm$	0.93	&	9.63 $\times$ 10$^{43}$	\\
HD 95089	&	nMS	&	2.17:1	&	496.57	$\pm$	2.88	&	20.65	$\pm$	0.55	&	2.40	$\pm$	0.05	&	228.83	$\pm$	2.88	&	6.73	$\pm$	0.57	&	4.64	$\pm$	0.11	&	-4.47	$\pm$	1.39	&	0.23	$\pm$	0.35	&	3.75 $\times$ 10$^{16}$	\\
11 Ursae Minoris	&	nMS	&	2.17:1	&	513.22	$\pm$	0.75	&	185.38	$\pm$	1.15	&	4.30	$\pm$	0.02	&	236.51	$\pm$	0.75	&	19.43	$\pm$	1.15	&	4.80	$\pm$	0.08	&	-11.14	$\pm$	2.18	&	-10.39	$\pm$	0.84	&	1.06 $\times$ 10$^{15}$	\\
HD 136418	&	nMS	&	2.17:1	&	474.99	$\pm$	1.26	&	41.62	$\pm$	0.39	&	5.63	$\pm$	0.03	&	218.89	$\pm$	1.26	&	10.41	$\pm$	0.47	&	2.86	$\pm$	0.08	&	-6.93	$\pm$	1.1	&	-4.99	$\pm$	0.42	&	8.03  $\times$ 10$^{12}$	\\
HD 81688	&	nMS	&	2.17:1	&	184.08	$\pm$	0.17	&	60.05	$\pm$	0.92	&	5.47	$\pm$	0.04	&	84.83	$\pm$	0.17	&	6.96	$\pm$	0.97	&	5.44	$\pm$	0.15	&	4.91	$\pm$	1.53	&	-1.21	$\pm$	0.67	&	1.41  $\times$ 10$^{11}$	\\
HIP 57050	&	MS	&	2.17:1	&	41.31	$\pm$	0.01	&	29.43	$\pm$	0.62	&	2.60	$\pm$	0.11	&	19.04	$\pm$	0.01	&	6.98	$\pm$	0.64	&	3.86	$\pm$	0.22	&	9.01	$\pm$	0.53	&	-15.77	$\pm$	0.83	&	4.27 $\times$ 10$^{9}$	\\
HD 206610	&	nMS	&	2.17:1	&	628.84	$\pm$	6.03	&	34.50	$\pm$	0.55	&	1.67	$\pm$	0.05	&	289.79	$\pm$	6.03	&	6.92	$\pm$	0.79	&	3.06	$\pm$	0.09	&	-17.3	$\pm$	1.7	&	19.30	$\pm$	0.47	&	1.59  $\times$ 10$^{7}$	\\
HD 216770	&	MS	&	2:1	&	121.77	$\pm$	0.38	&	14.05	$\pm$	2.97	&	4.30	$\pm$	0.33	&	60.88	$\pm$	0.38	&	33.86	$\pm$	3.44	&	0.79	$\pm$	0.18	&	36.91	$\pm$	6.87	&	31146.80	$\pm$	1.92	&	1.61  $\times$ 10$^{5}$	\\
HD 114386	&	MS	&	2.17:1	&	955.46	$\pm$	15.81	&	33.79	$\pm$	1.55	&	3.92	$\pm$	0.11	&	440.31	$\pm$	15.81	&	14.08	$\pm$	1.96	&	2.23	$\pm$	0.22	&	-5.24	$\pm$	3.82	&	33367.71	$\pm$	1.28	&	8.80  $\times$ 10$^{4}$	\\
o Coronae Borealis	&	nMS	&	2:1	&	187.61	$\pm$	0.13	&	33.64	$\pm$	0.67	&	6.00	$\pm$	0.05	&	93.81	$\pm$	0.13	&	8.64	$\pm$	0.67	&	4.66	$\pm$	0.10	&	1.92	$\pm$	0.57	&	1.82	$\pm$	0.49	&	8.66 $\times$ 10$^{4}$	\\
HD 101930	&	MS	&	2.17:1	&	71.30	$\pm$	0.17	&	17.93	$\pm$	0.41	&	1.38	$\pm$	0.04	&	32.86	$\pm$	0.17	&	2.37	$\pm$	0.50	&	1.49	$\pm$	0.20	&	-10.72	$\pm$	3.43	&	18363.19	$\pm$	0.35	&	3.00  $\times$ 10$^{4}$	\\
14 Andomedae	&	nMS	&	2.17:1	&	185.89	$\pm$	0.22	&	99.80	$\pm$	1.41	&	4.90	$\pm$	0.04	&	85.66	$\pm$	0.22	&	6.56	$\pm$	1.61	&	4.60	$\pm$	0.22	&	-7.64	$\pm$	2.96	&	1.68	$\pm$	1.24	&	2.59 $\times$ 10$^{4}$	\\
HD 180902	&	nMS	&	2.17:1	&	482.22	$\pm$	2.92	&	28.68	$\pm$	1.16	&	0.30	$\pm$	0.04	&	222.22	$\pm$	2.92	&	4.20	$\pm$	0.88	&	1.05	$\pm$	0.23	&	-3.74	$\pm$	2.69	&	9.27	$\pm$	0.54	&	1.83 $\times$ 10$^{4}$	\\
HD 218566	&	MS	&	2:1	&	225.54	$\pm$	0.14	&	7.60	$\pm$	0.24	&	0.90	$\pm$	0.07	&	112.77	$\pm$	0.14	&	2.47	$\pm$	0.26	&	0.17	$\pm$	0.17	&	0.61	$\pm$	0.11	&	0.82	$\pm$	0.19	&	1.70 $\times$ 10$^{4}$	\\
GJ 649	&	MS	&	2:1	&	601.38	$\pm$	2.17	&	11.55	$\pm$	0.31	&	3.41	$\pm$	0.11	&	300.69	$\pm$	2.17	&	2.88	$\pm$	0.39	&	1.83	$\pm$	0.32	&	0.86	$\pm$	0.31	&	6.18	$\pm$	0.46	&	1.10 $\times$ 10$^{4}$	\\
75 Ceti	&	nMS	&	2.17:1	&	694.41	$\pm$	1.40	&	37.12	$\pm$	0.74	&	3.84	$\pm$	0.04	&	320.01	$\pm$	1.40	&	3.31	$\pm$	0.70	&	4.32	$\pm$	0.21	&	4.59	$\pm$	0.48	&	0.40	$\pm$	0.47	&	1.04 $\times$ 10$^{4}$	\\
HD 27894	&	MS	&	2.17:1	&	17.97	$\pm$	0.01	&	58.40	$\pm$	0.49	&	5.04	$\pm$	0.07	&	8.28	$\pm$	0.01	&	3.91	$\pm$	0.73	&	0.50	$\pm$	0.17	&	-29.59	$\pm$	10.58	&	82907.65	$\pm$	1.94	&	6118.83	\\
HD 32518	&	nMS	&	2.17:1	&	157.45	$\pm$	0.19	&	117.90	$\pm$	2.19	&	6.00	$\pm$	0.03	&	72.56	$\pm$	0.19	&	9.90	$\pm$	2.38	&	4.92	$\pm$	0.17	&	13.84	$\pm$	4.14	&	-11.90	$\pm$	1.26	&	3259.36	\\
HD 231701	&	MS	&	2.17:1	&	141.30	$\pm$	0.35	&	41.56	$\pm$	1.48	&	1.32	$\pm$	0.10	&	65.11	$\pm$	0.35	&	12.81	$\pm$	3.37	&	2.98	$\pm$	0.23	&	-8.44	$\pm$	6.3	&	-0.02	$\pm$	1.81	&	993.37	\\
$\gamma_1$ Leonis	&	nMS	&	2:1	&	428.87	$\pm$	0.17	&	206.77	$\pm$	0.62	&	1.09	$\pm$	0.01	&	214.43	$\pm$	0.17	&	31.92	$\pm$	0.70	&	4.99	$\pm$	0.03	&	9.44	$\pm$	0.86	&	178.66	$\pm$	0.53	&	598.25	\\
HD 2638	&	MS	&	2.17:1	&	3.45	$\pm$	0.00	&	66.75	$\pm$	0.45	&	5.26	$\pm$	0.09	&	1.59	$\pm$	0.00	&	1.36	$\pm$	0.46	&	4.44	$\pm$	0.35	&	49.46	$\pm$	13.72	&	9619.32	$\pm$	2.32	&	407.32	\\
HD 31253	&	MS	&	2:1	&	464.44	$\pm$	0.64	&	10.75	$\pm$	0.34	&	0.27	$\pm$	0.06	&	232.22	$\pm$	0.64	&	3.58	$\pm$	0.33	&	2.87	$\pm$	0.13	&	0.59	$\pm$	0.17	&	1.96	$\pm$	0.25	&	295.22	\\
HD 221287	&	MS	&	2.17:1	&	452.51	$\pm$	1.00	&	73.50	$\pm$	1.37	&	2.10	$\pm$	0.04	&	208.53	$\pm$	1.00	&	12.95	$\pm$	2.46	&	2.34	$\pm$	0.05	&	10.72	$\pm$	2.36	&	-21861.09	$\pm$	1.29	&	124.62	\\
HD 190647	&	MS	&	2:1	&	931.10	$\pm$	76.66	&	30.25	$\pm$	2.44	&	3.81	$\pm$	0.44	&	465.55	$\pm$	76.66	&	7.90	$\pm$	1.04	&	2.94	$\pm$	0.93	&	-20.43	$\pm$	5.67	&	-40266.66	$\pm$	1.65	&	82.74	\\
HD 220773	&	MS	&	2:1	&	2877.77	$\pm$	87.74	&	14.60	$\pm$	1.51	&	1.63	$\pm$	0.12	&	1438.88	$\pm$	87.74	&	10.15	$\pm$	1.42	&	4.80	$\pm$	0.25	&	2.89	$\pm$	0.72	&	-4.96	$\pm$	1.05	&	54.06	\\
HD 330075	&	MS	&	2.17:1	&	3.39	$\pm$	0.00	&	106.81	$\pm$	0.73	&	5.87	$\pm$	0.01	&	1.56	$\pm$	0.00	&	0.49	$\pm$	0.49	&	3.22	$\pm$	1.21	&	-13.2	$\pm$	4.57	&	61278.58	$\pm$	0.40	&	37.21	\\
HIP 79431	&	MS	&	2.17:1	&	113.99	$\pm$	0.40	&	155.87	$\pm$	2.20	&	2.19	$\pm$	0.02	&	52.53	$\pm$	0.40	&	30.65	$\pm$	1.82	&	4.00	$\pm$	0.09	&	-385.58	$\pm$	24.21	&	10.53	$\pm$	1.90	&	26.99	\\
\hline							
\end{tabular}
}
\caption{The preferred fixed circular double models of the extended sample that has Bayes factor larger than the 95th percentile of the synthetic systems.  Under star type, MS and nMS refer to Main Sequence and non Main Sequence stars, respectively.  The order of the table is by Bayes factor.}
\label{tab:Results}
\end{center}
\end{table}
\twocolumn

Our primary analysis does not include stellar jitter (even though our synthetic data has jitter added to the simulated data).  We made this choice for a number of reasons.  One is that since we are considering a fixed period ratio, only noise that occurs at that specific frequency could produce a spurious signal.  Most sources of stellar noise occur on much different time-scales.  The stellar rotation periods (typically ranging from 4--40 days, \citep{McQuillan14}) are shorter than the inner planet periods for most of these systems.  Stellar p-mode oscillations have typical time-scales of 5-15 minutes \citep{Haywood15}.  And, surface granulation variations last minutes to hours, with the largest granules remaining on the surface of stars for about a day \citep{DelMoro04,Haywood15}.  The time-scales of long term stellar activity arising from the cyclical appearance of starspots are on the order of years to decades \citep{Strassmeier09}.

These facts support the interpretation that stellar noise is not the cause of the inner companion signal for the majority of our systems.  Nevertheless, we did a separate analysis that included a white noise jitter term to all models and found that \Nfixedwjitter\ of the \Nfixed\ systems still remain in the 95th percentile of likely two-planet systems, \NfixedwjitterTwoOneSeven\ of which prefer the 2.17:1 architecture.  Thus, even if we adopt the much more conservative approach---which assumes stellar jitter does indeed affect our data at precisely the relevant time scales---we still see a number of systems that favor the two-planet models.

While our results are primarily from the fixed circular double models, we examined the results of a floating circular double model in order to estimate the likely distribution of orbital periods for the inner companion.  We analyzed the real and synthetic systems with the floating circular double model and find an even larger portion of the systems that have Bayes factors above the 95th percentile---\Nfixedfloat\ systems, \Nfixedfloatpercent\ of the main sequence sample, (\Nfixedfloatextended\ systems, \Nfixedfloatextendedpercent, of the extended sample) with an estimated false positive rate of 0.95 $\pm$ 0.97 (2.1 $\pm$ 1.4 for the extended sample).  Of these \Nfixedfloat\ systems, \NfixedfloatFloating\ prefer the floating circular double model, \NfixedfloatTwoOneSeven\ prefer the fixed 2.17:1 model, and the remaining \NfixedfloatTwo\ prefer the 2:1 double circular model.  \Nfixedfloatwjitter\ systems remain in the 95th percentile when including stellar jitter in the model as a white noise term.

\begin{figure}
\includegraphics[width=\columnwidth]{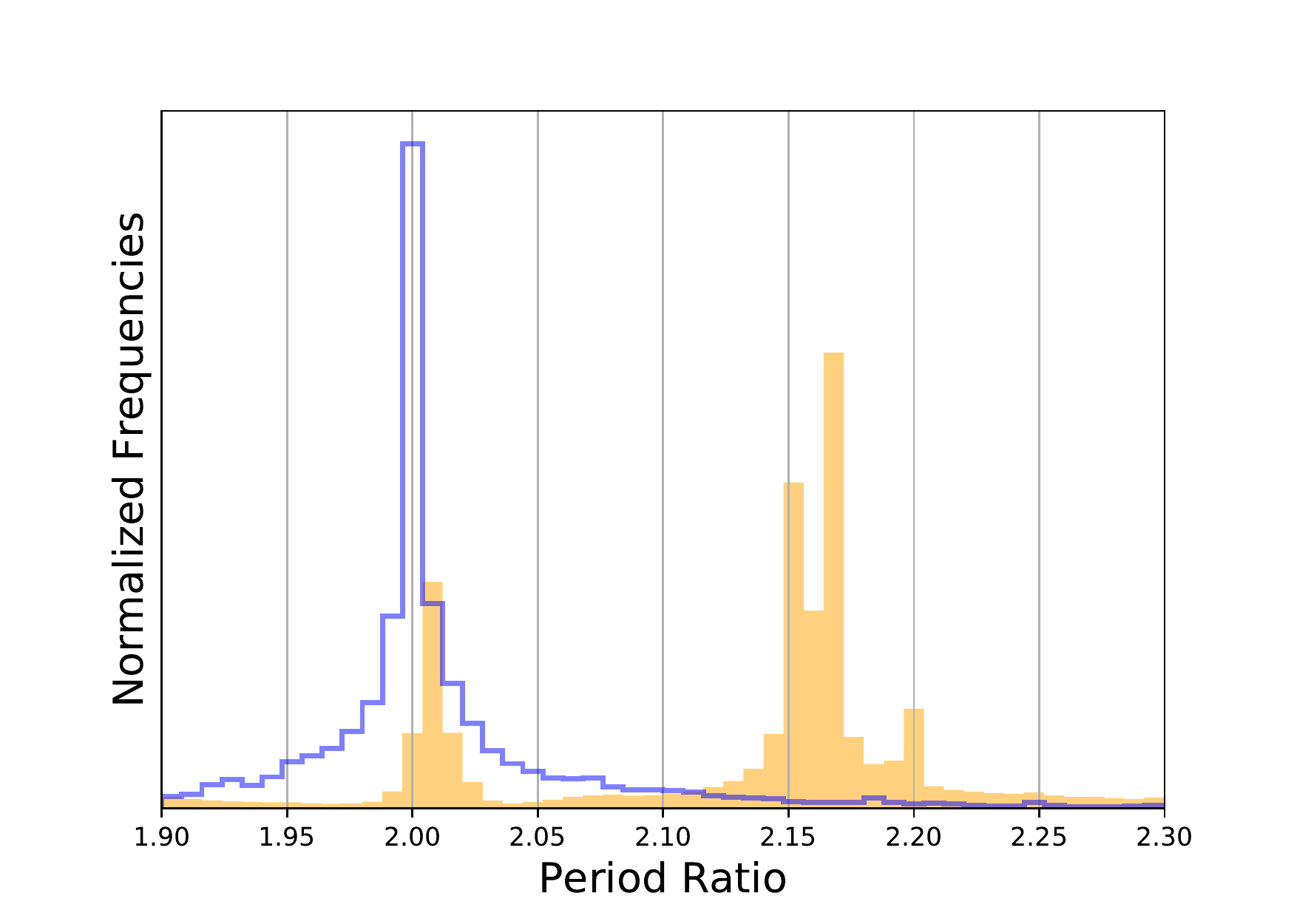}
\caption{The posterior distributions for the period ratio when considering the inner planet period as a free parameter.  The blue distribution is the 1,000 synthetic single eccentric systems.  This distribution peaks at 2:1---the location of the degeneracy.  The orange distribution is the 19 systems with Bayes factors larger than the 95th percentile thresholds that are hosted by main sequence stars.}
\label{fig:PRHistogramMS}
\end{figure}

We show the period ratio posteriors that result from this analysis for these \Nfixedfloat\ systems and the synthetic systems in Figure~\ref{fig:PRHistogramMS}.  These histograms show the combined, period ratio posterior distribution from fitting the circular double model without a constraint on orbital periods to the \Nfixedfloat\ systems and to the synthetic systems.  The distribution for the synthetic systems clearly shows the degeneracy at the location of the 2:1 MMR.  If the real systems (in orange) were single-planet systems, then the expected distribution should be the same as for the synthetics.  However, the two distributions differ significantly.  In fact, the distribution for the real systems mirrors the period ratio distribution from the \kepler\ data \citep{Steffen15}.  Most of the combined posteriors favor period ratios just wide of the 2:1 or between 2.15 and 2.2.  Only a few systems preferred the circular double model near the 2:1 because the degeneracy is located at the 2:1 and the power to distinguish between the models diminishes.  Thus, in that regime, more data with appropriate phase-sampling is essential to distinguish between the models.

\section{Discussion \& Conclusions} \label{sec:d&c}

There are currently 395 confirmed \EditTwo{solitary} RV planets\EditTwo{,} 
and we reanalyzed about fifteen percent of them.  Our extended sample contains nearly a quarter of the confirmed RV planets.  \Edit{The distribution of eccentricities, periods, velocity semi-amplitudes, etc. of our extended sample is shown in Figure~\ref{fig:parametercomparison}.}  If the \Nfixed\ systems in the main sequence sample (\Nfixedextended\ systems in the extended sample) that we identify are indeed circular doubles, then they would increase the number of RV multi-planet systems by $\sim$12.5\% ($\sim$25\%) since there are 120 confirmed systems reported with at least \Edit{two planets} discovered by RV.  They would also significantly alter the estimated mixture of these two architectures---shifting the relative importance of their implied dynamical histories.

\Edit{If the fraction of misidentified single eccentrics in the entire NASA Archive is similar to the misidentification fraction seen in our sample}, then there could be as many as $\sim$100 planets missing, or $\sim$15\% 
of the overall confirmed RV planets ($\sim$120 in the extended sample, or $\sim$18\% of the overall confirmed RV planets).    
Moreover, the apparent propensity for some systems to cluster around period ratios near 2.17 is a further indication that there is something fundamental, but still unknown, that attracts planet pairs into this period ratio.  We encourage observers to consider planning follow-up observations of these systems and make additional measurements at phases where the degeneracy is at its weakest.  New observations near these phases could confirm or refute the existence of these putative interior companions.  The success of such a campaign opens the door to identifying the architectures of the systems where the preferred model is still ambiguous.

\bibliographystyle{mnras}
\bibliography{mnrasbib}

\section*{Acknowledgements}
This research has made use of the NASA Exoplanet Archive, which is operated by the California Institute of Technology, under contract with the National Aeronautics and Space Administration under the Exoplanet Exploration Program. \linebreak http://exoplanetarchive.ipac.caltech.edu.  The authors declare no competing interests.  John H. Boisvert wrote the model fitting software, did the analysis, and wrote the paper.  Jason H. Steffen assisted with doing the analysis and editing the paper.  Benjamin E. Nelson wrote the analysis software.  We would like to thank Dan Foreman-Mackey for valuable conversations.  John H. Boisvert and Jason H. Steffen acknowledge support from NASA under grant NNH13ZDA001N-OSS issued through the Origins of Solar Systems program.  Benjamin E. Nelson acknowledges support from the Center for Interdisciplinary Exploration and Research in Astrophysics (CIERA) and the Data Science Initiative at Northwestern University.

\bsp  
\label{lastpage}
\end{document}